# Anisotropic Raman Scattering and Mobility in Monolayer $1T_d$-$ReS_2$ Controlled by Strain Engineering

Z. H. Zhou[1], B.C. Wei[1], Yimeng, Min[2], L. Z. Liu[1,a)],

[1] *Key Laboratory of Modern Acoustics, MOE, Institute of Acoustics and Collaborative Innovation Center of Advanced Microstructures, National Laboratory of Solid State Microstructures, Nanjing University, Nanjing 210093, P. R. China*

[2] *Kuang Yaming Honors School, Nanjing University, Nanjing 210093, P. R. China*

## Abstract

Regulation of electronic structure and mobility cut-on rate in two-dimensional transition metal dichalcogenides (TMDs) has attracted much attention because of its potential in electronic device design. The anisotropic Raman scattering and mobility cut-on rate of monolayer unique distorted-1T($1T_d$) $ReS_2$ with external strain are determined theoretically based on the density function theory. The angle-dependent Raman spectrum of $A_g$-like, $E_g$-like and $C_p$ models are used to discriminate and analysis structural anisotropy; the strain is exploited to adjust the structural symmetry and electronic structure of $ReS_2$ so as to enhance mobility cut-on rate to almost 6 times of the original value. Our results suggest the use of the strain engineering in high-quality semiconductor switch device.

**Keywords:** Anisotropy; $ReS_2$; Raman Scattering; Mobility; Strain Engineering

**PACS:** 73.20.At; 73.30.+y; 74.78.Fk

---

\* Corresponding author.   E-mail: lzliu@nju.edu.cn (L.Z.L.)

Recently, the transition metal dichalcogenides (TMDs) has attracted much attention because its tunable electronic and vibrational properties in promising applications.[1] They are expected to exhibit a wide range of electronic structures and exotic transport properties arising from the various electron configurations of transition metals, such assuperconductivity,[2] half-metallic magnetism[3] and charge density waves. Furthermore, as two-dimensional (2D) materials, they are the potential candidates to overcome the bottleneck: severe short-channel effects which conventional metal-oxide-semiconductor encountered.[4-10]

In this letter, we will be focused on a semiconductor material in TMDs family: rhenium disulfide ($ReS_2$). The distorted-1T ($1T_d$) $ReS_2$ takes on novel properties in optics, transport and vibration. A detailed analysis to previous researches disclose that the properties are closely related to anisotropy of the structure. Hong-Xia Zhong's group studied the anisotropic optical response and anisotropic excition of $ReS_2$ from PDOS analysis.[11] Erfu Liu's group measured the carrier mobility of the $ReS_2$ and revealed the transported property is different on various angles.[12] Daniel's group conducted experiments to obtain the angle-dependent Raman intensity distribution the modes ranging from 110 $cm^{-1}$ to 158 $cm^{-1}$ and noted that the relative intensities of the Raman modes are highly sensitive to the orientation of the in-plane crystallographic axes for few-layer samples.[13]

However, our understanding to essence the anisotropy of $ReS_2$ is so limited and hitherto no study has been done to adjust the anisotropy. External strain can be easily implemented by introducing a specific substrate in the fabrication of $T_d$ $ReS_2$ nanostructure. This is an effective way to regulate the structural anisotropy and carrier mobility and if the mobility cut-on rate can be adjusted, electronic device performance in fact be manipulated by strain engineering. So, in this letter, we will employ angle-resolved Raman spectrum to discriminate the anisotropy of the monolayer $T_d$ $ReS_2$ and apply external strain which is unavoidable especially in the frication of nanostructures to modify the carrier mobility anisotropy.[14]

The theoretical method is based on the density functional theory in Perdew-Burke-Ernzerhof (PBE) generalized approximation (GGA), using the CASTEP package code with projector augmented wave pseudopotenials.[15-17] The plane-wave energy cutoff of 750 eV is used to expand the Kohn-Sham wave functions and relaxation is carried out until convergence tolerances of 1.0 x 10$^{-5}$ eV for energy and 0.001Å for maximum displacement are reached. The Monkhorst-Pack *k*-point meshes (in two-dimensional Brillion zone) are 3x3x1 for the 2-D structure, which has been tested to converge.

Firstly, the structural symmetry are firstly display in Fig. 1. The 1x1 primitive cell of 1T$_d$ ReS$_2$ is shown in Fig. 1(a). The crystal lattice parameters we acquired from the calculation are listed as follow: OA= 6.57 Å, OB= 6.40 Å, ∠AOB= 60.16º, which is quite agreement with that of other groups.[1,12] We can see that the 1T$_d$ ReS$_2$ structure has asymmetric crystal structure, and its structural anisotropy is benefit to regulation of mobility cut-on rate. The Fig. 1(b) present the top view of the 3x3 supercell, the disparity between along a-axis and along b-axis (zigzag) is evident. The anisotropy can also be conveyed form the side view [see Fig. 1(c) and 1(d)]. Among the TMDs that have been reported to be stable individual layers of MoS$_2$, MoSe$_2$, WS$_2$, and WSe$_2$ have 1 hexagonal (1H) structure in their ground state and dichalcogens of Ti, V, and Ta are in the 1H phase.[1] However, the atomic structure of ReS$_2$ single layer is 1T$_d$ structure. One remarkable disparity between 1H, 1T and 1T$_d$ structure is that the symmetry and isomorphism around the principal axis. Hence the 1 T$_d$ structure of monolayer ReS$_2$ outstands the unique anisotropy atomic structure among TMDs family.

In order to discuss the anisotropic electron structure of 1 T$_d$ ReS$_2$, we make sections of the electron density distribution [see Fig. 2(a) slice 1 and slice 2]. The electron clouds of the S atoms located at almost the same height (S$_a$, S$_b$, S$_c$, S$_d$) are examined. The p-orbitals of S$_a$ and S$_b$ share little with each other [see slice 1], nevertheless, as it shown on slice 2. The S$_c$ and S$_d$ share electron so as to form an interaction parallel the OAB plane. Details of the S atoms around Re1 and the interactions exerted on Re1 are shown on Fig. 2 (b), (c) and (d), respectively. The maximum deviation rate ρ are defined as below to measure the anisotropy quantitatively:

$$\rho_f = \frac{f_{max} - f_{min}}{f_{min}} \qquad (1)$$

Where, $f$ represent Mulliken Charge of atoms (M), population of bonds(P), or length of bonds (L), respectively. $\rho_M = 850\%$, $\rho_P = 42\%$, $\rho_L = 8\%$. The anisotropy among the Mulliken Charge (effective charge) is most notable, which can also be inferred from the analysis to the slice 1 and slice 2. The anisotropy among the bond population is considerable either, which indicates that the iconicity and covalency of the bonds are various. The S6-Re1 bond is most covalent and S5-Re1 is most iconic. The variation among the length of the bonds partly explains the dissymmetry in lattice parameters.

Raman scattering is a useful tool to identify 2D material structural anisotropy, so the polarized-dependent Raman behavior are discussed subsequently. ReS$_2$ is isomorphic to the point group $C_i$,[13] the irreducible representations of the 36 Γ-point phonon modes (for primitive cell contains 12 atoms) can be written as

$$\Gamma = 18(A_g \oplus A_u). \qquad (2)$$

The 18 A$_g$ modes are Raman active and almost all of them can be detected in previous experiments.[13,18] Among the 18 A$_u$ modes, 15 are infrared active modes, and 3 are zero-frequency modes. As it is shown in the Table I, the frequency of 18 A$_g$ modes we obtained from GGA calculations are agree well with the theoretical and experiment results of other groups[18]. According to the classical Placzek approximation the anisotropic Raman intensity under the parallel configuration is[19,20]

$$I = |(\cos\theta, \sin\theta, 0)\tilde{R}(\cos\theta, \sin\theta, 0)^T|^2 \qquad (3)$$

where θ is the angle between the laser polarization (linear polarized) and zigzag direction.[19]; $\tilde{R}$ is the Raman tensor. We depict the angle-resolved Raman intensity distribution of several typical A$_g$-like, E$_g$-like and C$_p$ modes in Table II, with the vibration animation attached. In the C$_p$ mode at 368.1 cm$^{-1}$, the S atoms move horizontally; while in the A$_g$-like mode at 427.4 cm$^{-1}$ the S atoms vibrate vertically. So the vibration modes also outstand strong anisotropy. We also discover that the motion of the S atoms and Re atoms can be decoupled, for the C$_p$ mode at 270.3 cm$^{-1}$ shows that the S atoms shift while the Re atoms stay still. Scrutinizing the four angle-dependent intensity distribution (Table II), it is found that the direction where the intensity of the $C_p$ mode at 368.1 cm$^{-1}$ reach its maximum and the intensity of the $E_g$-like mode at 213.7 cm$^{-1}$ reach its minimum simultaneously is the zigzag direction. Thus, the diamond chain (zigzag) formed by Re atoms is recognized from the angle-dependent intensity distribution and the anisotropic atom structure can be determined.

Next, the angle-dependent carrier mobility is disclosed. The transport property of monolayer 1T$_d$ ReS$_2$ are governed to a large extent by carrier mobility. We apply a phonon-limited scattering model to calculation the mobility of the electron in conduction band minimum, in which the primary mechanism limiting carrier mobility is scattering due to phonons[21-24]

$$\mu_{2D} = \frac{e\hbar^3 C_{2D}}{k_B T m_x^* m_y^* (E_1^i)^2} \qquad (4)$$

Where $m_x^*$, $m_y^*$ are the effective mass in two principle axis [x (OC direction) and y (AB direction)] (the principle axis can is determined after calculation) of the effective mass ellipsoid. The $E_1^i$ represents thedeformation potential constant, defined by:

$$E_1^i = \Delta V_i / (\Delta l / l_0) \qquad (5)$$

where $\Delta V_i$ is the energy change of the i-th band under proper cellcompression and dilatation(usually 0.5% is suitable), $l_0$ is the lattice constant in the transport direction and $\Delta l$ is the deformation of $l_0$. The elastic modulus $C_{2D}$ is derived from:

$$C^{2D} = \frac{l_0^2}{S_0} \frac{\partial^2 E}{\partial^2 l} \qquad (6)$$

where $E$ is the energy of the primitive cell with area $S_0$. Temperature T is set as 300K. In the case no strain or compression exerted to the surface, our theoretical result and Erf Liu Group's experiment result[12] are shown in the Fig. 3(a). Both of them show that: the field-effect mobility is highly angle-dependent, with the largest value in the direction of 120° or 300°, which is 60° from the direction with the lowest value (180°). The anisotropic ratio of mobility $\frac{\mu_{max}}{\mu_{min}}$ is 3.1 (experiment[12]) and 5.9 (our calculation), which is noticeably larger than other reported 2D anisotropic materials, such as 1.8 for thin-layer black phosphorus.[25] Our theoretical result depicts the butterfly-like anisotropy mobility profile which is well resemble to what experiment result do. However, the deviation exist and the reason should be discussed. First and foremost, the sample may be impure so the scatter caused by foreign substance cannot be ignored.

Secondly, the structure was suffered from deformation when measured which accounts for deviation.

The Anisotropy of the mobility can be explicated by the scattering mechanism. The probability that electron is elastically scattered from $|\varphi_{n\mathbf{k}}>$ to $|\varphi_{n\mathbf{k'}}>$ is written as below:

$$W(n\mathbf{k}, n\mathbf{k'}) = \frac{2\pi}{\hbar} |<\varphi_{n\mathbf{k'}}|\Delta V|\varphi_{n\mathbf{k}}>|^2 \, \delta(\varepsilon_{n\mathbf{k'}} - \varepsilon_{n\mathbf{k}}) \qquad (7)$$

Where $\Delta V$ is the perturbing potential due to lattice vibration. According to the deformation potential theory:[26]

$$\Delta V(\mathbf{r}) = D_{ac}\Delta(\mathbf{r}) \qquad (8)$$

where $\Delta(r) = \Delta\Omega/\Omega_0$ is the Relative variation of the volume and the $D_{ac}$ is the deformation potential constant. $\Delta(r) = \frac{\Delta\Omega}{\Omega_0}$ is isotropy, while $D_{ac}$ is anisotropic, thus according to formula (8) the $\Delta V(\mathbf{r})$ is anisotropy and so the scattering probability is various with the angle[see formula (7)]. For larger scattering probability which means larger disturbance to electron will lead to smaller mobility, the mobility variation with angle can be comprehended.

As an innovative job, we propose a method that employing the strain to modify the configuration of the lattice to achieve larger anisotropic ratio of mobility. As the mobility at 120° direction is the maximum under no compression and strain, so it is likely to get stronger anisotropy by raising the mobility at 120°. The mobility at 120° as a function of single axis strain is show in Fig. 3(b). The mobility rises to peak at -3% then declines under strain applied along a-axis while the mobility goes down continuously under strain applied along b-axis. The anisotropy correspond to the -3% strain along a-axis (OA=6.31Å) is worth attention. We calculate the anisotropy under lattice parameter a(OA) =b(OB) = 6.31Å and ∠AOB= 60°[see Fig.3(c)] and discover the $\mu_{max}/\mu_{min}$ is raised up to 33.6, almost 6 times of the theoretical value in Fig.3(a). It is also found that in the case a(OA) =b(OB) = 6.57 Å and ∠AOB= 60°, the anisotropy feature a main peak at 120° a smaller peak at 0°, and the minimum at 60°, so the tristate control is possible via this structure[see Fig. 3(d)].

In conclusion, anisotropic atomic and electronic structure were demonstrated and discriminated by angle-dependent Raman scattering behavior. For more prominent swift and control apparatus, the method applying strain to enhance the mobility anisotropy is proposed and proved to be feasible theoretically. A very distinctive direction: 120º direction can be regarded as a strain transformation invariant in isotropy, for the carrier mobility at 120º remains the largest under the strain. Now, we can expound the mechanism according to the deformation potential theory (see formula 7 and 8): the perturbing potential is least sensitive to the strain or compression at 120º, thus the scatter probability is minor than other directions hence leads to the most significant mobility. The anisotropic response of band structure to the lattice vibration is the nature of the transport anisotropy. Since the external strain can be easily applied by means of specific substrate in fabrication, the strategy to apply external strain in engineering of mobility cut-on rate in electronic device has large potential.

**ACKNOWLEDGMENTS**

This work was supported by National Basic Research Programs of China under Grants Nos. 2014CB339800 and 2013CB932901 and National Natural Science Foundation (No. 11374141 and 11404162). Partial support was from Natural Science Foundations of Jiangsu Province (No. BK20130549).We also thank the computational resources provided by High Performance Computing Centre of Nanjing University and High Performance Computing Centre of Shenzhen.

**Table captions**

**Table I.** Our calculation result, Feng Group's calculation result[18] and experiment result[18] of 18 $A_g$ modes.

**Table II.** Angle-dependent Raman intensity and corresponding vibration animations of $E_g$-like mode at 213.4 cm$^{-1}$, $C_p$ mode at 270.3 cm$^{-1}$, $C_p$ mode at 368.1 cm$^{-1}$ and $A_g$-like mode 427.4 cm$^{-1}$.

| Symmetry | Ourresult (cm$^{-1}$) | Calculation[18] (cm$^{-1}$) | Expt[18] (cm$^{-1}$) |
|---|---|---|---|
| A$_g$-like | 134.7 | 129.3 | 139.2 |
| A$_g$-like | 138.2 | 137.2 | 145.3 |
| E$_g$-like | 151.5 | 148.3 | 153.6 |
| E$_g$-like | 161.5 | 158.4 | 163.6 |
| E$_g$-like | 213.4 | 208.9 | 217.7 |
| E$_g$-like | 220 | 228.7 | 237.7 |
| C$_p$ | 258.2 | 261.9 | 278.3 |
| C$_p$ | 261.3 | 268.8 | 284.7 |
| E$_g$-like | 299.5 | 295.9 | 307.8 |
| E$_g$-like | 301.7 | 298.2 | 311.0 |
| C$_p$ | 306.4 | 303.5 | 320.6 |
| C$_p$ | 312.1 | 311.1 | 324.9 |
| C$_p$ | 336.5 | 332.7 | 348.8 |
| C$_p$ | 359.0 | 354.7 | 369.5 |
| C$_p$ | 368.1 | 363.3 | 377.4 |
| C$_p$ | 398.7 | 393.5 | 408.3 |
| A$_g$-like | 411.0 | 406.8 | 419.3 |
| A$_g$-like | 427.6 | 424.7 | 437.5 |

Zhou *et al.*   Table I.

| Raman frequency Mode | Angle-dependent Raman intensity | Vibration animation |
|---|---|---|
| 213.4 cm$^{-1}$ $E_g$-like | | |
| 270.3 cm$^{-1}$ $C_p$ | | |
| 368.1 cm$^{-1}$ $C_p$ | | |
| 427.4 cm$^{-1}$ $A_g$-like | | |

Zhou *et al.* Table II.

**Figure caption**

Fig. 1 (a) 1x1 cell of 1 layer ReS$_2$ with $1T_d$ structure. z-axis is perpendicular to the paper. (b) 3x3 supercell include the zigzag diamond chain lined in Red. (c) The side elevation of the 3x3 supercell projected along the a-axis. (d) The side elevation of the 3x3 supercell projected along the b-axis.

Fig. 2 (a) Two slices of electron density distribution. Slice 1: the slice plane contain Re1 and keep parallel to z axis and OC. Slice 2: the slice plane contain Re1 and keep parallel to z axis and OB. (b) Mulliken Charge of the atoms S1, S2, S3, S4, S5, S6. (c) and (d) Length and Population of the bonds S1-Re1 (Bond1), S2-Re1 (Bond2), S3-Re1 (Bond3), S4-Re1 (Bond4), S5-Re1 (Bond5), S6-Re1 (Bond6).

Fig. 3 (a) Theoretical result and experiment result[12] of the mobility depended on angle. The a-axis and b-axis are correspond to the 0° and 60.16° respectively. (b) The mobility in 120° change when single strain is applied to a-axis or b-axis. Polynomial fit (order =3) is applied. (c) Theoretical result of the mobility anisotropy when lattice parameters are confined to a (OA)=b(OB)= 6.31 Å and ∠AOB=60°. The a-axis and b-axis are correspond to the 0° and 60° respectively. (d) Theoretical result of the mobility anisotropy when lattice parameters are confined to a (OA)=b(OB)= 6.57 Å and ∠AOB=60°. The a-axis and b-axis are correspond to the 0° and 60° respectively.

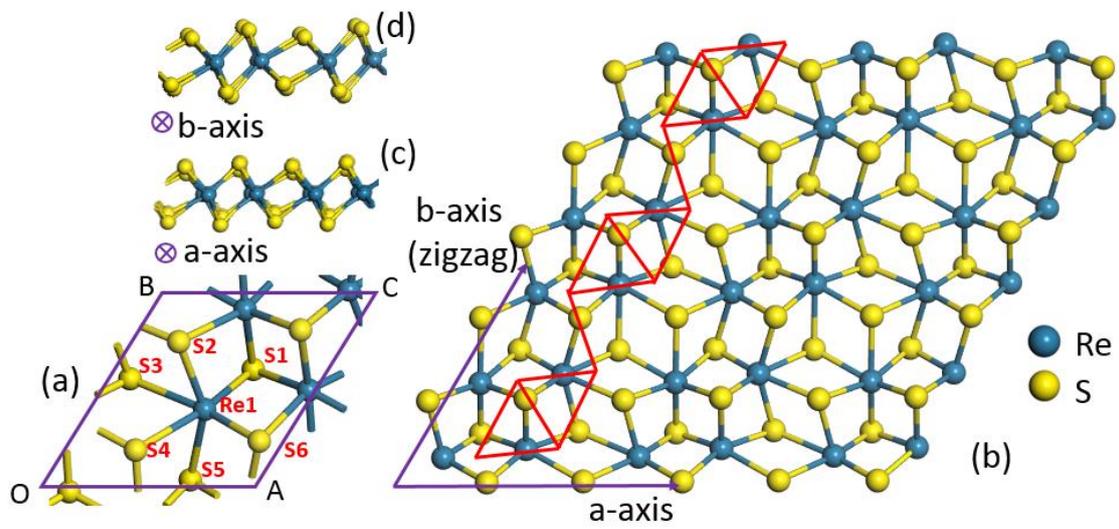

FIG. 1, Zhou. *et al*.

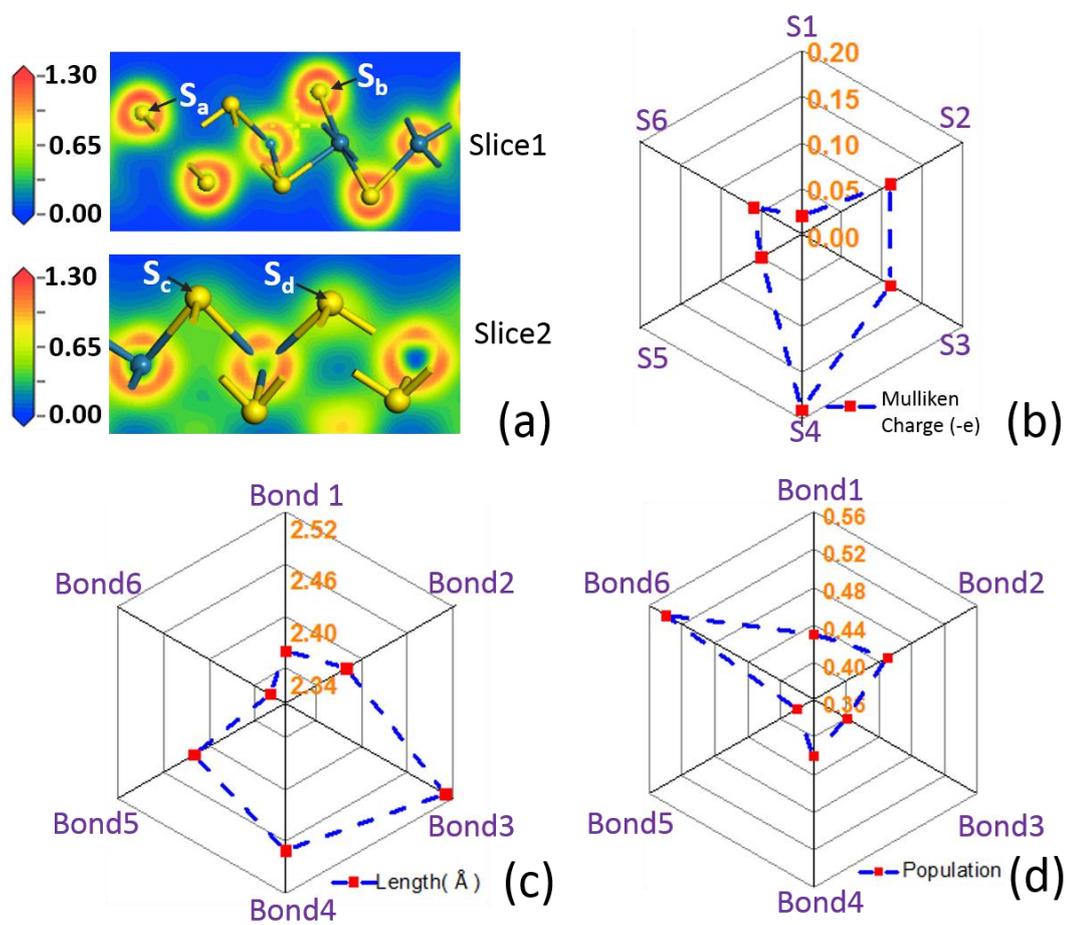

FIG. 2, Zhou. *et al*.

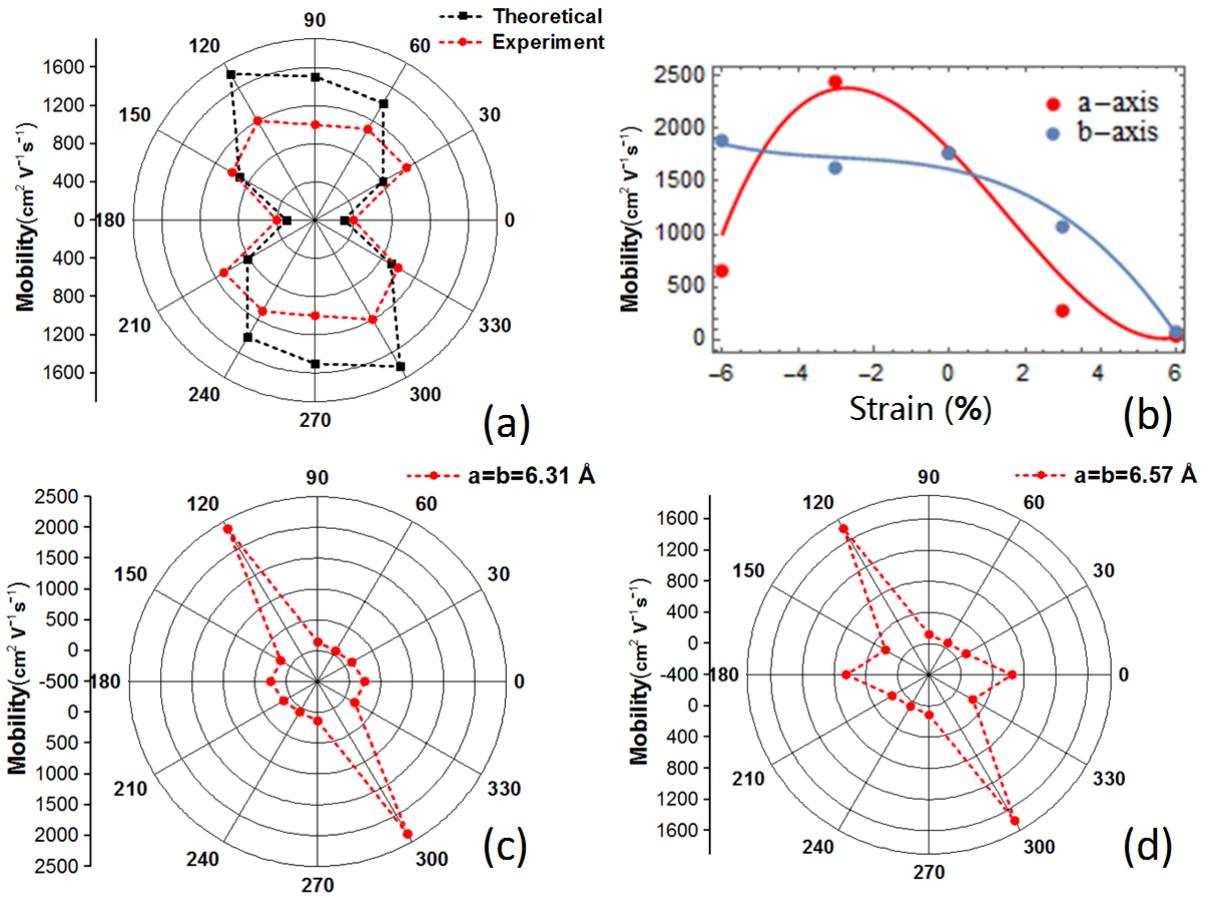

FIG. 3, Zhou. *et al*.

16